\documentclass[a4paper,12pt]{article}
\pdfoutput=1
\usepackage{graphicx, rotating,colortbl}
\usepackage{hyperref}
\usepackage{slashed}
\usepackage{ifpdf}
\usepackage{charter}
\usepackage[dvipsnames]{xcolor}

\def\bea{\begin{eqnarray}}
\def\eea{\end{eqnarray}}

\ifx\pdfoutput\undefined
\usepackage[dvips,bookmarks=false]{hyperref}	% This is for arXiv.org
\else
\usepackage{hyperref}	% This is for pdftex
\fi
\hypersetup{colorlinks,bookmarksopen,bookmarksnumbered,citecolor=verdes,
linkcolor=blus,pdfstartview=FitH,urlcolor=rossos}
\def\hhref#1{\href{http://arxiv.org/abs/#1}{#1}} % in bibliography
\usepackage{amsfonts}
\usepackage{amsmath}
\usepackage{slashed}

\newcommand{\beq}{\begin{equation}}
\newcommand{\eeq}{\end{equation}}
\newcommand{\fig}[1]{~\ref{fig:#1}}

\newcount\Mac  \Mac=1  % devo mettere Mac=1 se sto lavorando sul file Mac
\newcommand{\ifMac}[2]{\ifnum\Mac=1 #1 \else #2 \fi}
\def\putps(#1,#2)(#3,#4)#5#6{\ifnum\Mac=1 \put(#1,#2){\special{picture #5}}
\else  \put(#3,#4){\includegraphics{#6}} \fi}

\newcommand{\One}{\hbox{1\kern-.24em I}}

\newcommand{\GeV}{\,{\rm GeV}}
\newcommand{\TeV}{\,{\rm TeV}}

\newcommand{\eV}{\,{\rm eV}}
\newcommand{\NP}{Nucl. Phys.}

\newcommand{\eq}[1]{~{\rm (\ref{eq:#1})}}

\newcommand{\lascia}[1]{}
\makeatletter
\def\art{\@ifnextchar[{\eart}{\oart}}
\def\eart[#1]#2#3#4#5#6{{\rm #2}, {#3 #4} {\rm (#6) #5} [arXiv:{\hhref{#1}}]}
\def\hepart[#1]#2{{\rm #2, arXiv:\hhref{#1}}}
\newcommand{\oart}[5]{{\rm #1}, {#2 #3} {\rm (#5) #4}}

\newcounter{alphaequation}[equation]
\def\thealphaequation{\theequation\hbox to
0.6em{\hfil\alph{alphaequation}\hfil}}
\def\eqnsystem#1{
\def\@eqnnum{{\rm (\thealphaequation)}}
\def\@@eqncr{\let\@tempa\relax \ifcase\@eqcnt \def\@tempa{& & &} \or
  \def\@tempa{& &}\or \def\@tempa{&}\fi\@tempa
  \if@eqnsw\@eqnnum\refstepcounter{alphaequation}\fi
\global\@eqnswtrue\global\@eqcnt=0\cr}
\refstepcounter{equation} \let\@currentlabel\theequation \def\@tempb{#1}
\ifx\@tempb\empty\else\label{#1}\fi
\refstepcounter{alphaequation}
\let\@currentlabel\thealphaequation
\global\@eqnswtrue\global\@eqcnt=0 \tabskip\@centering\let\\=\@eqncr
$$\halign to \displaywidth\bgroup \@eqnsel\hskip\@centering
$\displaystyle\tabskip\z@{##}$&\global\@eqcnt\@ne
\hskip2\arraycolsep\hfil${##}$\hfil& \global\@eqcnt\tw@\hskip2\arraycolsep
$\displaystyle\tabskip\z@{##}$\hfil
\tabskip\@centering&\llap{##}\tabskip\z@\cr}

\def\endeqnsystem{\@@eqncr\egroup$$\global\@ignoretrue} \makeatother

\def\Lag{{\cal L}}
\def\SU{{\rm SU}}

\def\circa#1{\,\raise.3ex\hbox{$#1$\kern-.75em\lower1ex\hbox{$\sim$}}\,}

\usepackage{multicol}
\usepackage{color}
\definecolor{rosso}{cmyk}{0,1,1,0.4}
\definecolor{rossos}{cmyk}{0,1,1,0.55}
\definecolor{rossoc}{cmyk}{0,1,1,0.2}
\definecolor{blu}{cmyk}{1,1,0,0.3}
\definecolor{blus}{cmyk}{1,1,0,0.6}
\definecolor{bluc}{cmyk}{1,1,0,0.1}
\definecolor{verde}{cmyk}{0.92,0,0.59,0.25}
\definecolor{verdec}{cmyk}{0.92,0,0.59,0.15}
\definecolor{verdes}{cmyk}{0.92,0,0.59,0.4}
\definecolor{grigio}{cmyk}{0,0,0,0.07}
\definecolor{rosa}{cmyk}{0,0.1,0.1,0.02}
\definecolor{rosino}{cmyk}{0,0.05,0.05,0.02}
\definecolor{rosas}{cmyk}{0,0.3,0.25,0.05}
\definecolor{celeste}{cmyk}{0.1,0,0,0.02}
\definecolor{giallino}{cmyk}{0,0,0.4,0.02}
\definecolor{rosso}{cmyk}{0,1,1,0.4}
\definecolor{rossos}{cmyk}{0,1,1,0.55}
\definecolor{rossoc}{cmyk}{0,1,1,0.2}
\definecolor{blu}{cmyk}{1,1,0,0.3}
\definecolor{bluc}{cmyk}{1,1,0,0.1}
\definecolor{blucc}{cmyk}{0.7,0.5,0,0}
\definecolor{viola}{cmyk}{0,1,0,0.6}
\definecolor{viola2}{cmyk}{0,1,0.2,0.6}
\definecolor{verde}{cmyk}{0.92,0,0.59,0.25}
\definecolor{verdec}{cmyk}{0.92,0,0.59,0.15}
\definecolor{verdes}{cmyk}{0.92,0,0.59,0.4}
\definecolor{verdino}{cmyk}{0.12,0,0.09,0.05}
\definecolor{giallo}{cmyk}{0,0,1,0}
\definecolor{gialloverde}{cmyk}{0.44,0,0.74,0}

\font\tenrsfs=rsfs10 at 12pt

\font\sevenrsfs=rsfs7
\font\fiversfs=rsfs5
\newfam\rsfsfam
\textfont\rsfsfam=\tenrsfs
\scriptfont\rsfsfam=\sevenrsfs
\scriptscriptfont\rsfsfam=\fiversfs
\def\mathscr#1{{\fam\rsfsfam\relax#1}}
\def\Lag{\mathscr{L}}

\begin{document}
% IFUP-TH/2011-1\hfill CERN-PH-TH/2010-XX
\color{black}
%\vspace{0.5cm}
\begin{center}
{\Huge\bf\color{Magenta} A modified
naturalness principle\\
and its experimental tests
}
\bigskip\color{black}\vspace{0.3cm} \\[3mm]
{{\large\bf  Marco Farina$^{a}$, Duccio Pappadopulo$^{b,c}$
{\rm and}  Alessandro Strumia$^{d,e}$}
} \\[7mm]
{\it  (a)  Department of Physics, LEPP, Cornell University, Ithaca, NY 14853, USA}\\
{\it  (b) Department of Physics, University of California, Berkeley, CA 94720, USA}\\
{\it  (c) Theoretical Physics Group, Lawrence Berkeley National Laboratory, Berkeley, USA}\\
{\it  (d) Dipartimento di Fisica dell'Universit{\`a} di Pisa and INFN, Italia}\\
{\it  (e) National Institute of Chemical Physics and Biophysics, Tallinn, Estonia}\\
\end{center}
\bigskip

\centerline{\large\bf\color{blus} Abstract}
\begin{quote} %\large
Motivated by LHC results, we modify the usual criterion for naturalness  by ignoring the uncomputable power divergences.
The Standard Model satisfies the modified criterion (`finite naturalness')
for the measured values of its parameters.
Extensions of the SM motivated by observations
 (Dark Matter, neutrino masses, the strong CP problem, vacuum instability, inflation)
satisfy finite naturalness in special ranges of their parameter spaces which often
imply new particles below a few TeV.
Finite naturalness bounds are weaker than usual naturalness bounds because
any new particle with SM gauge interactions gives a finite contribution to the Higgs mass at two loop order.
\color{black}
\end{quote}
\tableofcontents
\newpage

\section{Introduction}
The naturalness principle strongly influenced high-energy physics in the past decades~\cite{thooft}, leading to the belief
that physics beyond the Standard Model (SM)  must exist at a scale $\Lambda_{\rm NP}$ such that quadratically divergent
quantum corrections to the Higgs squared mass are  made finite
(presumably up to a log divergence)
and not much larger than the Higgs mass $M_h$ itself.
This ideology started to conflict with data
after TeVatron measured the top mass (which implies a sizeable order-one top Yukawa coupling $\lambda_t$)
and after LEP excluded new charged particles below 100 GeV~\cite{FT}.
Indeed, imposing that the SM one loop correction to $M_h^2$
\beq \delta m_h^2 \approx\delta m_h^2 ({\rm top~loop})\approx \frac{12\lambda_t^2}{(4\pi)^2} \Lambda^2_{\rm NP}
\times\left\{\begin{array}{l}
1 \cr
\ln M_{\rm Pl}^2/\Lambda_{\rm NP}^2
\end{array}\right.\,
\label{nat} \eeq
is smaller than
$ M_h^2\times\Delta$, where
$\Delta$ is the amount of allowed
fine-tuning (ideally $\Delta\circa{<} 1$), implies\footnote{Computing the fine tuning with respect to electroweak parameters as in~\cite{EWFT}
corresponds to the first possibility.  
The second possibility corresponds to
computing the fine tuning with respect to high-scale parameters.
}
\beq \Lambda_{\rm NP}\circa{<} \sqrt{\Delta}\times
\left\{\begin{array}{l} 400\GeV \\ 50\GeV
\end{array}\right. \ .
 \label{eq:stop}\eeq
The most plausible new physics motivated by naturalness is supersymmetry.  It adds new particles at the weak scale,
with the lightest one possibly being Dark Matter (DM).
Further support for this scenario come from  gauge unification and from the fact that
DM loosely around the weak scale  is  indicated
by the hypothesis  that DM is the thermal relic of a massive stable particle.
Doubting that nature is natural seemed impossible.

%In many models the quadratic divergence is tamed into a log divergence,
%such that eq.\eq{stop} is enhanced by a $\ln M_{\rm Pl}^2/m_h^2$, which brings
%$\Lambda_{\rm NP}\circa{<} 50\GeV\times\sqrt{\Delta}$.

\medskip

However, no new physics has been so far seen at LHC with $\sqrt{s}=8\TeV$,
such that, in models that aim to be valid up to high energies,
the  unit of measure for $\Delta$ presently is
the kilo-fine-tuning.
While this is not  conclusive evidence, 
while special models that minimise fine-tuning are being considered,
while naturalness arguments can be weakened by allowing
for a finer tuning, while various searches have not yet been performed,
while LHC will run at higher energy, etc,
it is fair to say that the most straightforward interpretation of present data
is that the naturalness ideology is wrong.

\medskip

This situation lead to consider the opposite extremum:
the Higgs is
light due to huge cancellations~\cite{ant} because  `anthropic selection'
destroyed naturalness.

%justified by invoking `anthropic selection'.
%If there is no new physics at the TeV scale, the axion becomes the most plausible DM candidate.

\medskip

Here we explore an intermediate possibility, that sometimes appeared in the literature,
more or less implicitly.
We name it {\em `finite naturalness'}.
The idea is that we should ignore the
uncomputable quadratic divergences, so that the Higgs mass is
naturally small as long as there are no heavier particles that give large finite contributions to the Higgs mass.

\medskip

Whether or not quadratic divergences are really present is decided by the cut-off of the theory.
Various condensed matter systems exhibit spontaneous symmetry breaking phenomena like the SM:
experiments show that light scalars like the Higgs are not natural in that context, where the
the cut-off is given by atomic physics which indeed behaves like lattice regularisation,
with power divergences cut-offed by the lattice size.

On the other hand, the cut-off of the SM is unknown.  
Very likely it is relativistically invariant so that it is not a lattice.
Some authors explored the possibility that quadratic  divergences vanish
because the true physical cut-off behaves like dimensional regularization.
For example, this might maybe happen in special quantum gravity models where the Planck scale arises from
the spontaneous breaking of a dilatation-like symmetry; the weak scale
can dynamically arise as the scale where the quartic coupling of some extra singlet scalar becomes negative~\cite{Shap}.
Another possibility is an infinite tower of states at the Planck scale, arranged in a way that cancels power divergences~\cite{Dienes}.
Alternatively, new physics might allow for a Veltman throat at the Planck scale (a possibility which has been excluded
within the SM~\cite{SMextrapolated,SMextrapolated2}).
%Another possibility is that the quantum gravity scale might be  lowered down to the weak scale in the context of extra dimensional models.
These speculations are not  based on theoretically firm grounds.

\medskip

Anyhow, the goal of this paper is not advocating for the `finite naturalness' scenario.

Instead, we want to explore how
experiments can test if it satisfied in nature.

For example the SM satisfies finite naturalness (as discussed in section~\ref{SM}), and high-scale gauge unification does not satisfy it.
While gauge unification is just a nice theoretical hypothesis, there are kinds of new physics which are demanded
by experiments: neutrino masses (discussed in section~\ref{nu}), Dark Matter
(discussed in section~\ref{DM}), and possibly the strong CP problem (section~\ref{CP}) and vacuum stability and inflation (section~\ref{vac}).
We will classify models of new physics from the point of view of finite naturalness,
finding constraints on their parameter spaces which usually imply new particles not much above the weak scale.
Conclusions are given in section~\ref{concl}.

\section{The Standard Model}\label{SM}
Here and in the rest of the paper we write the SM Higgs potential at tree level as
\beq V = - \frac{m^2}{2} |H|^2 + \lambda |H|^4\eeq
such that,  expanding around the minimum of the Higgs doublet
$ H=(0,(v+h)/\sqrt{2})$ with
$v = m/\sqrt{2\lambda}\approx 246.2\GeV$,
the parameter $m$ at tree level equals the physical Higgs mass,
$M_h = \sqrt{2\lambda} v=m$.

\begin{figure}[t]
\begin{center}
$$\includegraphics[width=0.45\textwidth]{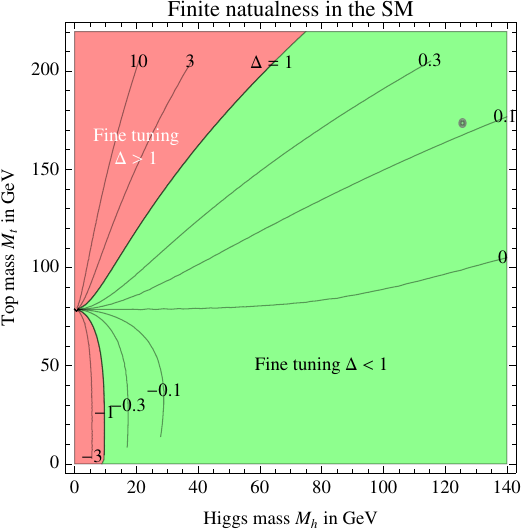}$$
\caption{\em The SM satisfies `finite naturalness' for the observed values of its parameters
(small ellipse), while a large fine-tuning would be present for a lighter Higgs.
\label{fig:SMnat}}
\end{center}
\end{figure}

The expectation is that finite naturalness is satisfied by the SM quantum corrections, because the top and the vectors are not much heavier than the Higgs.
To confirm this we compute at one-loop accuracy
the Higgs mass parameter $m$, in the $\overline{\rm MS}$ scheme.
We find
\beq \label{eq:mm}
m^2  = M_h^2 + \hbox{Re}\,\Pi_{hh}^{(1)}(p^2=M_h^2)|_{\rm finite} + 3\frac{T^{(1)}|_{\rm finite}}{v}\eeq
where $M_h$ is the pole Higgs mass,
$\Pi_{hh}^{(1)}(p^2=M_h^2)$ is the on-shell one-particle-irreducible Higgs propagator
and  $T^{(1)}$ is the Higgs tadpole.
Their sum reconstructs the full Higgs propagator.
Computing them in a generic $R_\xi$ gauge
we find that their combination in eq.\eq{mm} is gauge-independent, as it should~\cite{SM-NLO}, and explicitly given by\footnote{
Equivalent expressions for the one loop SM correction to the pole Higgs mass have already been presented in various papers,
including appendix~C of~\cite{KJ},
eq.~(3.14) of~\cite{Martin}, appendix A of~\cite{SMextrapolated2}.
These computations are here used (maybe for the first time)
to extract the fundamental SM parameter $m$ from data with one-loop accuracy;
the result with two-loop accuracy will appear in~\cite{SM-NLO}.}
 \bea
m^2 &=&M_h^2 +
\frac{1}{(4\pi v)^2} \bigg[  6M_t^2(M_h^2-4M_t^2)B_0(M_h;M_t,M_t)+ 24 M_t^2 A_0(M_t) +\nonumber\\
&& +(M_h^4-4M_h^2M_W^2+12M_W^4)B_0(M_h;M_W,M_W)-2(M_h^2+ 6M_W^2) A_0(M_W) +  \nonumber  \\
&&+\frac{1}{2}\left(M_h^4-4 M_h^2 M_Z^2+12 M_Z^4\right)B_0(M_h;M_Z,M_Z) -(M_h^2+ 6M_Z^2) A_0(M_Z) +  \nonumber  \\
&&+\frac{9}{2} M_h^4 B_0(M_h;M_h,M_h)-3M_h^2 A_0(M_h) \bigg] \\
&=& M_h^2 \bigg(1+0.133+ \beta_m^{\rm SM} \ln\frac{\bar\mu^2}{M_t^2}\bigg)
\label{eq:d1Mh}
\eea
where
\beq A_0 (M) = M^2(1-\ln\frac{M^2}{\bar\mu^2}),\qquad
B_0(p;M,M) = -\int_0^1 \ln\frac{M^2-x(1-x)p^2}{\bar\mu^2}dx\eeq
are the finite parts of the usual Passarino-Veltman functions and
$M_t$ is the top quark mass, $M_W$ is the $W$ mass, $M_Z$ is the $Z$ mass.
This correction reproduces the well known one-loop SM RGE equation for $m^2$
\beq
\frac{d m^2}{d\ln\bar\mu^2} = \beta_m^{\rm SM}  m^2,\qquad
\beta_m^{\rm SM}=\frac{3}{4}\frac{4 y_t^2+8\lambda-3g^2-g_Y^2}{(4\pi)^2}.\eeq
In view of the log divergence, the finite part of the correction to $m^2$ is scheme-dependent;
it depends on the value of $\bar\mu$ and on its definition (e.g.\ 
when choosing  $\overline{\rm MS}$ instead then MS).\footnote{In general, the constant terms here and in the following computations
depend on the regulator as well as an on the renormalisation scheme;
we will use $\overline{\rm MS}$.  On the other hand,
the log-enhanced terms are univocally defined and correspond to the coefficients of the
usual renormalisation group equations.}
From eq.\eq{d1Mh} we see that the $\overline{\rm MS}$ Higgs mass parameter equals
$m(\bar\mu=M_t) =132.8\GeV$.
Renormalizing it at large energies~\cite{RGE3loop} we find  $m(\bar\mu = M_{\rm Pl}) = 141.1\GeV$.
As a consequence the SM satisfies `finite naturalness', for the observed values of its parameters.
Fig.\fig{SMnat} shows contour-levels of the fine-tuning
$\Delta \equiv m^2(M_t)/M_h^2-1$: we see
that $\Delta\approx 0.13$ is small for the observed values of the SM parameters, while
a Higgs mass $\approx 10$ times lighter than the top  would have led to a
`finite naturalness' problem within the SM.

%At one loop, one has $m(m)\approx 1.06 M_h$ for $\overline{\rm MS}$ parameter $m$ renormalised at the scale $m$.

\medskip

%As discussed in the introduction,
%the potential conflict between the SM and finite naturalness concerns gravitational quantum corrections to the Higgs mass,
%presumably of order $M_{\rm Pl}^2$.
We now explore the implications for `finite naturalness' of new physics
motivated by observations.
%neutrino masses, by the observed Dark Matter,  by the strong CP problem and by SM

\section{Finite naturalness, neutrino masses and leptogenesis}\label{nu}
The observation of neutrino masses~\cite{review}, presumably of Majorana type, points to new physics at some scale
possibly as high as $v^2/m_\nu \sim 10^{14}\GeV$.  At tree level, neutrino masses can be mediated by 3 types of new particles,
called type I, II and III see-saw.
We will study the corrections to the Higgs mass parameter in these scenarios.

\subsection{Type-I see saw}
Type-I see-saw contains heavy right-handed neutrinos $N$ with mass $M$ and Yukawa couplings $\lambda_N~NLH$
to lepton doublets $L$
such that at low energy one obtains Majorana neutrino masses $m_\nu = \lambda_N^2 \langle H\rangle^2/M$~\cite{review}.
The one-loop correction to the squared Higgs mass is
\beq \delta m^2 = \frac{4\lambda_N^2}{(4\pi)^2} M^2 ( \ln \frac{M^2}{\bar\mu^2}-1)\label{eq:typeImm}
 \eeq
where the RGE scale $\bar\mu$ can be identified with the cut-off of the theory, possibly the Planck scale.
Indeed the log-enhanced term in eq.\eq{typeImm}
means that, above $M$, heavy right-handed neutrinos produce the following term
\beq \frac{d m^2}{d\ln\bar\mu^2} = \frac{4 \lambda_N^2}{(4\pi)^2 } M^2 + \beta_m^{\rm SM} m^2\eeq
in the RGE for the Higgs mass term squared parameter $m^2$.
Therefore the condition of finite naturaless, $\delta m_h^2 \circa{<}M_h^2\times \Delta$
(where $\Delta$ is an order-one fine-tuning factor),
is satisfied by type-I models if right-handed neutrinos are lighter than~\cite{Vissani}
\beq
M \circa{<} M_h \left(\Delta \frac{16\pi^2 M_h}{m_\nu}\right)^{1/3} \approx 0.7~10^7\GeV\times\sqrt[3]{\Delta}
\qquad\hbox{(Type-I see-saw)}
\label{eq:typeIM}\eeq
having assumed $m_\nu =(\Delta m^2_{\rm atm})^{1/2}\approx 0.05\eV$.
This upper bound on $M$ is hardly compatible with thermal leptogenesis,
that needs $M\circa{>}2~10^9\GeV$, unless right-handed neutrinos
dominated the energy density of the universe~\cite{leptog} (such that leptogenesis
can be successful for  $M\circa{>} 2\cdot 10^7\GeV$)
and/or in presence of resonant enhancements.

\medskip

It is interesting to discuss what happens if condition\eq{typeIM} is not satisfied.  Then, $m^2$ must have at large energy a large positive value (that corresponds to a more unstable potential)
in order to get the observed small $M_h\approx 125\GeV$ at low energy.
Given that  the quartic coupling $\lambda$ becomes $\lambda\approx 0$ when renormalised at a scale
around $10^7\GeV$, the one-loop potential might there develop new features\footnote{We thank J.R. Espinosa
and J.E. Mir\'o
for having raised this issue.}
 (such as a new local minimum).
We verified that this is not the case.

%...we checked that at   $h\sim M$ the full one-loop potential, given by
%\beq V^{(1)} = -\sum_i \frac{M_i^4}{2(4\pi)^2}(\ln\frac{M_i^2}{\bar\mu_i^2} - \frac{3}{2}),\eeq
%(where the sum runs over the neutrino mass eigenstates)
%does not develop any particular structure (such that a new local minimum).

\subsection{Type-III see saw}
In type-III see-saw models the heavy singlets $N$ are replaced by heavy weak triplets $N^a$ with zero hypercharge~\cite{review}.
Beyond a correction to $m^2$ similar to eq.\eq{typeImm}, there is now a bigger correction induced at two loops
by the SU(2)$_L$ gauge couplings of the triplet.
Using the results for the generic two-loop potential~\cite{V2loop} we obtain
%\footnote{
%The term proportional to $M^2$ in the effective potential at 2 loops is~\cite{V2loop}
%\beq V = \frac{g_2^2 }{2(4\pi)^4} \Tr(T^a T^a)  (
%f_{FF V}(M^2,M^2,M_V^2)-M^2 f_{\bar F\bar F V}(M^2,M^2,M_V^2))\eeq
%Next
%$$ f_{FF V}(M^2,M^2,M_V^2) \stackrel{M\gg M_V}{\simeq} M^2 M_V^2 (13 -6\ln \frac{M_V^2}{M^2}\ln\frac{M^2}{\bar\mu^2}-
%3\ln^2\frac{M^2}{\bar\mu^2}) $$
%$$ f_{\bar F\bar F V}(M^2,M^2,M_V^2) \stackrel{M\gg M_V}{\simeq} M_V^2 (11+ 12 \ln\frac{M^2}{\bar\mu^2}-6\ln \frac{M_V^2}{M^2} \ln\frac{M^2}{\bar\mu^2} - 3\ln^2 \frac{M^2}{\bar\mu^2} )$$
%so the $\ln^2 \bar\mu^2$ disappears as it should:
%\beq V  \stackrel{M\gg M_V}{\simeq}
%\frac{g_2^2 }{2(4\pi)^4} \Tr(T^a T^a)  M^2   M_V^2 (2 -12\ln\frac{M^2}{\bar\mu^2})
%\eeq
%next using $M_V^2 = g_2^2 v^2/2$ and
%$${\rm Tr}\, T^a T^b = \delta^{ab}\frac{n}{12}(n^2-1)=\{0,2,10\}$$
%one gets for a $n$-dimesional representation of SU(2)
%\beq \delta m^2  =  - \frac{g_2^4}{2(4\pi)^4}\frac{n(n^2-1)}{4} M^2 (2 -12\ln\frac{M^2}{\bar\mu^2})
%\label{eq:dmmtypeIII}\eeq
%and for $n=3$ the coefficient of the ln is the same as the one coming from RGE.}
\beq \delta m^2 = \frac{g_2^4}{(4\pi)^4} M^2 (36\ln\frac{M^2}{\bar\mu^2}-6).
\label{eq:dmmtypeIII}\eeq
The log-enhanced term reproduces the RGE equation for $m^2$
(that  we computed at 2 loops using~\cite{RGE2loop}):
\beq \frac{d m^2}{d\ln\bar\mu^2} =  \frac{36 g_2^4}{(4\pi)^4 } M^2 + \beta_m^{\rm SM} m^2.
\eeq
%\beq \delta m_h^2 \approx\frac{g^4}{(4\pi)^4} M^2\circa{<}m_h^2. \eeq
Assuming that the cut-off of the theory is around the Planck mass we find that the condition of
finite naturalness, $\delta m_h^2 \circa{<}M_h^2\times \Delta$,  demands that the weak triplet must be lighter than
\beq M  \circa{<}0.94 \TeV\times\sqrt{\Delta}\qquad\hbox{(Type-III see-saw)} .\eeq
Such a low mass is not compatible with successful thermal leptogenesis in the context of
type-III  see-saw~\cite{leptogIII,Sommerfeld}, but can lead to observable signals at LHC~\cite{typeIIILHC}.

\subsection{Type-II see saw}
 Type II see-saw employs one scalar  $T^a$,
triplet under $\SU(2)_L$ and  with hypercharge $Y=1$~\cite{review}.  Its relevant couplings are
 \beq \label{eq:Ltriplet}
\Lag = \Lag_{\rm SM} +|D_\mu T|^2 - M^2 |T|^2 +\frac{1}{2}
({\lambda}_T^{ij} L^i  \varepsilon \tau^a L^j T^a  +  \lambda_H M \, ~H  \varepsilon \tau^a H ~T^{a*} +\hbox{h.c.})
-\lambda_{HT} |H|^2 |T|^2+\cdots
\eeq
where $\lambda_T$ is a symmetric flavour matrix,
$\varepsilon$ is the permutation matrix,
and $\tau^a$ are the usual $\SU(2)_L$ Pauli matrices.
Majorana neutrino masses arise as
$m_\nu= {\lambda}_T \lambda_H \langle H\rangle^2/M^2$.
We ignore the corrections to $m^2$ due to
non-minimal couplings non  related to neutrino masses, except for
$\lambda_{HT}$, because it will enter the subsequent discussion.

At one loop there is  the  correction to $m^2$ induced by $\lambda_H$:
\beq \delta m^2 = -\frac{6\lambda_H^2 M^2}{(4\pi)^2} ( \ln\frac{M^2}{\bar\mu^2} -1).\label{eq:dmmIInu}
\eeq
At two loops there is the correction induced only by electroweak interactions:
%\footnote{
%Using the generic results of Martin, the two-loop potential for $g_Y=0$ is
%\beq V = \frac{3 g_2^2 }{(4\pi)^4}  M^2 M_V^2 (3 +12\ln\frac{M_V^2}{M^2}+20\ln\frac{M^2}{\bar\mu^2}-12\ln \frac{M_V^2}{M^2}\ln\frac{M^2}{\bar\mu^2}-
%9\ln^2\frac{M^2}{\bar\mu^2})\eeq
%which means an apparently wrong result
%\beq \delta m^2 =-\frac{3 g_2^4 }{(4\pi)^4}  M^2 (3 +12\ln\frac{M_V^2}{M^2}-16\ln\frac{M^2}{\bar\mu^2}-12\ln \frac{M_V^2}{M^2}\ln\frac{M^2}{\bar\mu^2}-
%9\ln^2\frac{M^2}{\bar\mu^2})
%\label{eq:dmmII}
%\eeq
%Changing a sign in Martin results gives instead
% \beq
% V = \frac{3 g_2^2 }{(4\pi)^4}  M^2 M_V^2(7+4\ln\frac{M^2}{\bar\mu^2}+3\ln^2\frac{M^2}{\bar\mu^2})
% \label{eq:dmmII}
% \eeq
% and the $\delta m^2$ in the text.
%%For a generic scalar multiplet the general expression should be
%%\beq V = \frac{ g_2^2 }{2(4\pi)^4} \Tr(T^a T^a)  M^2 M_V^2 (3 +12\ln\frac{M_V^2}{M^2}+20\ln\frac{M^2}{\bar\mu^2}-12\ln \frac{M_V^2}{M^2}\ln\frac{M^2}{\bar\mu^2}-
%%9\ln^2\frac{M^2}{\bar\mu^2})\eeq
%}
%\xxx{or (why fattore 2 diversi?)(dove sta $g_Y$ nella prima? come fa a venire somma di $g^4$? $M_V=M_W$?}
 \beq\label{eq:dmmII}
 \delta m^2=-M^2\frac{6 g_2^4 +3 g_Y^4}{(4\pi)^4} (\frac{3}{2}\ln^2\frac{M^2}{\bar\mu^2}+2 \ln\frac{M^2}{\bar\mu^2}+\frac{7}{2}).
 \eeq
%\beq \delta m^2 \simeq \frac{9 M^2 }{2}\frac{6g_2^4 + 3 g_Y^4 }{(4\pi)^4}\ln^2 \frac{M^2}{\bar\mu^2}\eeq
The $\ln^2\bar\mu$ term in eq.\eq{dmmII} arises as a composition of two one-loop effects:
i) the coupling $\lambda_{HT}$ is generated  from pure gauge effects;
ii)  $\lambda_{HT}$ affects the Higgs mass.
These two effects are  described by the following terms in the one loop RGE equations (which had previously been computed in~\cite{RHEtypeII}):
\beq \frac{d\lambda_{HT}}{d\ln\bar\mu} =   \frac{6 g_2^4+ 3g_Y^4}{(4\pi)^2}+\cdots \qquad
\frac{d  m^2}{d\ln\bar\mu} =-  \frac{12 \lambda_{HT}}{(4\pi)^2} M^2+\cdots \eeq
The $\ln\bar\mu$ term in eq.\eq{dmmII} arises from the finite parts in these one-loop effects, combined with the pure
two-loop RGE for $m^2$ %(that we compute from eq. (3.23) of~\cite{RGE2loop})
\beq \frac{dm^2}{d\ln\bar\mu} = -10\frac{6 g_2^4 +3 g_Y^4}{(4\pi)^4} M^2 .\eeq
% So 1 and 2 loop RGE mean that
%\beq \delta m^2 \simeq -M^2
%\frac{6g_2^4 + 3 g_Y^4 }{(4\pi)^4}
%\left(
% \frac{3  }{2}\ln^2 \frac{M^2}{\bar\mu^2}+
%5 \ln \frac{M^2}{\bar\mu^2}\right)
%\eeq
Assuming $\bar\mu\sim M_{\rm Pl}$,
the two loop gauge correction to $m^2$ of eq.\eq{dmmII} imply the finite naturalness bound
\beq M\circa{<} 200\GeV\times\sqrt{\Delta}\qquad\hbox{(Type-II see-saw)}\eeq	
which is comparable with present LHC bounds~\cite{typeIILHC}.
Then the one-loop correction of eq.\eq{dmmIInu} sets a weak upper bound on $\lambda_H$, which
does not forbid the possibility that $T\to HH$ is the dominant decay channel.
Leptogenesis is not possible at such a low value of $M$~\cite{leptogII}.

\medskip

In conclusion, `finite naturalness' and neutrino masses suggest that
type III and especially type II see-saw allow for direct signals at LHC.
However, the most plausible possibility seems type-I see-saw
with leptogenesis around $10^7\GeV$.

\section{Finite naturalness and Dark Matter}\label{DM}
Dark Matter exists, and presumably it is some new fundamental particle.
However, its mass $M$ is totally unknown, ranging from the galactic scale to the Planck scale.
The hypothesis that DM is a thermal relic suggests that $M$ might be around the weak scale.
Even so, there is a huge number of viable DM models. Here we focus on some representative models
that, unlike supersymmetric DM, are not motivated by the usual formulation of the hierarchy problem.
We choose Minimal Dark Matter~\cite{MDM} as a representative of models where DM has electro-weak interactions,
and the minimal scalar singlet~\cite{scalarsinglet} and fermion singlet~\cite{fermionsinglet}
models as representatives of models where DM has no electro-weak interactions.
Axion DM is considered in the subsequent axion section.

\subsection{Minimal Dark Matter}
Minimal Dark Matter (MDM) assumes that DM is the neutral component of just one electroweak multiplet
which only has electroweak gauge interactions.
As summarized in table~\ref{tab:MDM},
a few choices of the multiplets are possible, and in each case everything is predicted as function of the DM mass $M$,
which can be fixed assuming that the thermal relic density equals the observed DM density~\cite{MDM}.

Coming to the condition of finite naturalness, we compute the correction to the Higgs squared mass term
which is produced by electroweak gauge interactions at two loops.

\begin{itemize}
\item For a generic fermionic multiplet with hypercharge $Y$ and dimension $n$ under $\SU(2)_L$ we find
%\footnote{
%For a generic multiplet with hypercharge $Y$ and dimension $n$ under $\SU(2)_L$,
%the two loop diagrams are proportional to
%$$\Tr(T^A T^B)(\tau^A \cdot \tau^B)_{ij}= n (\frac{n^2-1}{4} g_2^4 + Y^2 g_Y^4) \delta_{ij}$$
% where $A$ runs over all 4 EW vectors, $T$ are the multiplet generators and $\tau$ the Higgs doublet generators,
% with gauge couplings inside them.
%  We used, for SU(2),
%$${\rm Tr}\, T^a T^b = \delta^{ab}\frac{n}{12}(n^2-1)=\{0,\frac{1}{2},2,5,10,\ldots\}\qquad
%(\tau^a\tau^a)_{ij} = 3 \delta_{ij}$$}
\beq \delta m^2 =\frac{cn M^2}{(4\pi)^4}  \bigg(\frac{n^2-1}{4} g_2^4 +  Y^2 g_Y^4\bigg)
\bigg(6\ln\frac{M^2}{\bar\mu^2}-1\bigg)
\label{eq:dmmFermion}
\eeq
where $c=1$ for Majorana fermions ($Y=0$ and odd $n$) and $c=2$ for Dirac fermions ($Y\neq 0$ and/or even $n$).
For $n=3$ and $Y=0$ we recover the type-III see-saw result of eq.\eq{dmmtypeIII}.

\begin{table}[t]
\begin{center}
\small
$$\begin{array}{|ccc|cccccc|}\hline
\multicolumn{3}{|c|}{\hbox{Quantum numbers}}
&\hbox{DM could}&\!\!\!\hbox{DM mass}&m_{{\rm DM}^\pm} - m_{\rm DM}\!\!\!&
\hbox{Finite naturalness} & \hbox{$\sigma_{\rm SI}$ in}  \\
 \SU(2)_L &\! {\rm U}(1)_Y\! &\hbox{Spin} &
 \hbox{decay into} & \hbox{in TeV} &\hbox{in MeV} & \hbox{bound in TeV} &
 \hbox{$10^{-46}\,{\rm cm}^2$} \\ \hline
 \hline \rowcolor[rgb]{0.95,0.95,0.95}
%1 & 0 & 0 & HH^*& ?&-&0&?\\ \rowcolor[rgb]{0.95,0.95,0.95}
%1 & 0 & 1/2 & LH^* & - &-&0&0\\  \rowcolor[rgb]{0.95,0.95,0.95}
%\hline
2 & 1/2 & 0 & EL & 0.54  & 350 & 0.4 \times\sqrt{\Delta}& (0.4\pm 0.6)\,10^{-3} \\  \rowcolor[rgb]{0.95,0.95,0.95}
2 & 1/2 & 1/2 & EH & 1.1  & 341 & 1.9\times\sqrt{\Delta} & (0.3\pm 0.6)\,10^{-3}\\  \rowcolor[cmyk]{0.32,0,0.19,0.01}
\hline
3 & 0 & 0 & HH^* & 2.0\to2.5  & 166 & 0.22\times\sqrt{\Delta}& 0.12\pm 0.03\\  \rowcolor[cmyk]{0.32,0,0.19,0.01}
3 & 0 & 1/2 & LH & 2.4\to 2.7  & 166 & 1.0\times\sqrt{\Delta} & 0.12\pm 0.03\\  \rowcolor[rgb]{0.95,0.95,0.95}
3 & 1 & 0 & HH,LL & 1.6\to \,?  & 540 & 0.22 \times\sqrt{\Delta}&0.001\pm 0.001\\ \rowcolor[rgb]{0.95,0.95,0.95}
3 & 1 & 1/2 & LH & 1.9\to\, ? & 526 & 1.0\times\sqrt{\Delta} & 0.001\pm 0.001\\  \rowcolor[rgb]{0.95,0.95,0.95}
\hline
4 & 1/2 & 0 & HHH^* & 2.4\to\,?  & 353 & 0.14\times\sqrt{\Delta} & 0.27\pm 0.08\\  \rowcolor[rgb]{0.95,0.95,0.95}
4 & 1/2 & 1/2 & (LHH^*) & 2.4\to\,?  & 347 & 0.6\times\sqrt{\Delta} & 0.27\pm 0.08\\  \rowcolor[rgb]{0.95,0.95,0.95}
4 & 3/2 & 0 & HHH & 2.9\to\,? & 729 & 0.14\times\sqrt{\Delta} & 0.15\pm 0.07\\  \rowcolor[rgb]{0.95,0.95,0.95}
4 & 3/2 & 1/2 & (LHH) & 2.6\to\, ?  & 712 & 0.6\times\sqrt{\Delta} &0.15\pm 0.07\\ \rowcolor[rgb]{0.95,0.95,0.95}
\hline\rowcolor[rgb]{0.95,0.95,0.95}
5 & 0 & 0 & (HHH^*H^*) &5.0 \to 9.4  & 166 & 0.10\times\sqrt{\Delta} & 1.0\pm0.2\\   \rowcolor{green}
5 & 0 & 1/2 & \hbox{stable} &4.4 \to 10  & 166 & 0.4\times\sqrt{\Delta} & 1.0\pm0.2\\  \rowcolor[cmyk]{0.32,0,0.19,0.01}
\hline
7 & 0 & 0 & \hbox{stable} & 8\to 25 & 166 & 0.06\times\sqrt{\Delta} & 4\pm 1\\
\hline
\end{array}
$$
\color{black}
\normalsize
\caption{\em {\bf Minimal Dark Matter}. The first columns define the quantum numbers of the possible DM weak multiplets.
Next we show the possible decay channels which need to be forbidden;
the DM mass predicted from thermal abundance (the arrows indicate the effect of taking into
account non-perturbative Sommerfeld corrections,
which have not been computed in all cases);
the predicted splitting between the charged and the neutral components of the DM weak multiplet;
the bound from finite naturalness and the prediction for the Spin-Independent
direct detection cross section on protons $\sigma_{\rm SI}$.
\label{tab:MDM}}
\end{center}
\end{table}%

\item For a scalar multiplet we find
\beq \delta m^2 =-\frac{n M^2}{(4\pi)^4}  \bigg(\frac{n^2-1}{4} g_2^4 +  Y^2 g_Y^4\bigg)
\bigg(\frac{3}{2}\ln^2\frac{M^2}{\bar\mu^2}+2 \ln\frac{M^2}{\bar\mu^2}+\frac{7}{2}\bigg). \label{eq:MDMS}\eeq
%\xxx{no factor $c=2$ here for complex multiplets because massive scalars do not need the conjugate representation?}
For $n=3$ and $Y=0$ we recover the type-II see-saw result of eq.\eq{dmmII}.
\end{itemize}

\medskip

We then show in table~\ref{tab:MDM} the finite naturalness upper bounds on $M$ for the various possible MDM multiplets.
Furthermore,  table~\ref{tab:MDM} shows the predictions for the DM mass $M$ suggested by the hypothesis
that DM is a thermal relic with cosmological abundance
\beq\Omega_{\rm DM}h^2 = 0.1187\pm 0.0017~\cite{Planck}.\eeq
(Such results differ from the analogous table of~\cite{MDM} because $M$ has been recomputed taking into account
Sommerfeld effects~\cite{MDMsommerfeld}, which lead to the change indicated by the arrows in table~\ref{tab:MDM}).
We see that for $n\circa{>}4$ finite naturalness is not compatible with the larger value of $M$ suggested by thermal DM.
We recall that the most motivated MDM candidate is a fermion with $n=5$ (or a scalar with
$n=7$) because such particles are automatically stable: just like in the case of the proton the first possible source
of DM decay comes from dimension-6 operators, and thereby are naturally suppressed by a large scale.

\medskip

In table~\ref{tab:MDM} we also recomputed the prediction for the spin-independent direct detection cross section $\sigma_{\rm SI}$.
With respect to the analogous table of~\cite{MDM} we recomputed $\sigma_{\rm SI}$ taking into account:
a) the correct value of the Higgs mass, which has now been measured;
b) two-loop DM/gluon interactions which partially cancel the one-loop DM/quark interactions~\cite{Hisano};
c) improved determinations of the nucleon matrix elements.
In particular new experimental measurements of $\pi p$ scattering
and new lattice simulations point to a strange content of the nucleon
$f_s \equiv \langle N|m_s \bar s s|N\rangle/m_N = 0.043\pm0.011$ \cite{walkerloud},
lower than the value $f_s\sim 0.3$ previously assumed.
All these effect reduce the predicted value of $\sigma_{\rm SI}$. We refer to table~1 of \cite{Hisano2} for all other nuclear inputs.
Following~\cite{MDM}, for the scalar MDM candidates we neglected the
contribution to $\sigma_{\rm SI}$ from a possible quartic coupling between DM and the Higgs.
Furthermore, pure MDM candidates with $Y\neq 0$ are excluded by a
too large tree level contribution to $\sigma_{\rm SI}$ from $Z$ exchange.
We assumed that such effect is removed by adding a small mass mixing
between the real and imaginary component of the neutral component, such that
DM is a real particle and only the loop contribution is present~\cite{MDM}.

%{\em R [http://arxiv.org/pdf/1211.4873.pdf]}

\medskip

The MDM weak multiplets present at low energy change the running of $g_2,g_Y$ such that
the SM couplings renormalised at large energy can satisfy the Veltman throat condition at the Planck scale
(this is roughly achieved by adding a doublet or a $Y=0$ triplet)
or give absolute stability of the Higgs potential (this needs larger corrections to the $\beta$-functions).

\begin{figure}
\begin{center}
\includegraphics[width=0.55\textwidth]{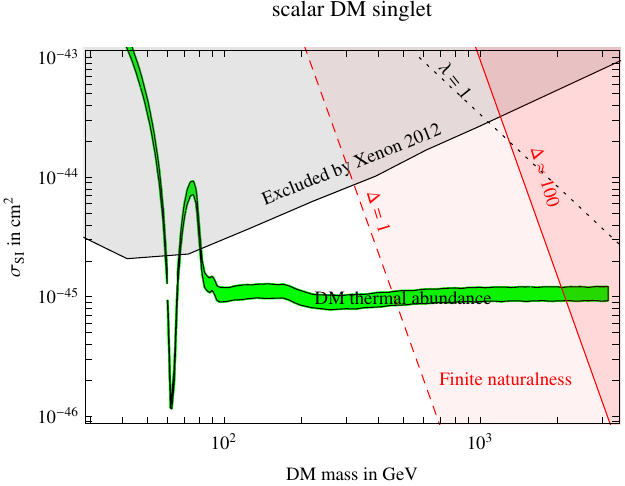}
\caption{\em  {\bf Scalar singlet DM model}.
In the ($M,\sigma_{\rm SI}$) plane we plot the region excluded by the Xenon direct search experiment,
the band favored by the thermal DM abundance, the upper bound on the DM mass following from
finite naturalness for $\Delta=1$ and $\Delta=100$.
\label{fig:SS}}
\end{center}
\end{figure}

\subsection{The scalar singlet DM model}\label{scalarsinglet}
We now consider DM without SM gauge interactions.
A scalar singlet $S$ that respects a $S\to -S$ symmetry  is added to the SM, such that $S$ is a stable DM candidate.
The model is described by the Lagrangian~\cite{scalarsinglet}
\beq \Lag =\Lag_{\rm SM } + \frac{(\partial_\mu S)^2}{2} - \frac{m_S^2}{2}S^2
-\lambda_{HS} S^2 |H|^2 - \frac{\lambda_S}{4} S^4\  .
\label{eq:L}\eeq
The DM mass is given by $M^2 = m_S^2 + 2\lambda_{HS} v^2$.
The quantum correction to the Higgs mass parameter $m^2$ is
\beq \delta m^2 =- \frac{2\lambda_{HS}^2}{(4\pi)^2} M^2 ( \ln \frac{M^2}{\bar\mu^2}-1)\label{eq:scalarsinglet}
\qquad\hbox{i.e.}\qquad
%%\beq \delta m^2 \approx \frac{\lambda_SH}}{(4\pi)^2} M^2\eeq
 \frac{d m^2}{d\ln\bar\mu^2} =  -\frac{2\lambda_{HS}^2}{(4\pi)^2 } M^2+\beta_m^{\rm SM} m^2
\eeq
such that the finite naturalness bound on the DM mass depends on the unknown Higgs/DM coupling $\lambda_{HS}$.
This coupling can be fixed by DM physics.
Indeed, the spin-independent  cross section relevant for direct DM detection is predicted as~\cite{scalarsinglet}
\beq\label{eq:f} \sigma_{\rm SI}
= \frac{\lambda_{HS}^2 m_N^4 f^2}{\pi M^2 M_h^4} \eeq
where $f\approx 0.295$ is the nucleon matrix element.
In fig.\fig{SS} we plot the finite naturalness bound in the ($M,\sigma_{\rm SI}$) plane.
Imposing $\delta m^2 \circa{<} M_h^2\times\Delta $ for a running from the Planck scale down to the weak scale we find
that no fine-tuning needs DM  to be lighter than a few hundred GeV (red dashed curve), which corresponds to a coupling $\lambda_{HS}\circa{<}0.1$.
By either allowing for a $\Delta \approx 100$ fine-tuning, or by dropping the
$\ln (M_{\rm Pl}^2/M^2)\approx 80$ RGE enhancement in the loop effect,
gives  the weaker
bound represented by the continuous red curve.
This bound is stronger than the one obtained by demanding perturbativity of $\lambda_{HS}$
(exemplified by the dotted curve that corresponds to $\lambda_{HS}=1$).

\smallskip

Furthermore, in the same plot we show the region (green band) where $\lambda_{HS}$ is fixed by the assumption that the thermal relic DM density
reproduces the measured DM density $\Omega_{\rm DM}$,
as computed in tree-level approximation~\cite{scalarsinglet}.
%\duccio{Sommerfeld enhancement at large masses is negligible?}
%\duccio{Perturbativit\`a di $\lambda$ fino a $\Lambda~ 10^8$\,GeV, $m_{DM}<5$\,TeV}

We incidentally observe that quadratic divergences to $m^2$ could vanish because of a
Veltman throat at the Planck scale: this requires $\lambda_{HS}(M_{\rm Pl})\approx 0.2$
having assumed the SM values for all other couplings.

\subsection{The fermion singlet model}
We here consider a model where DM is a fermionic singlet $\psi$ with no SM gauge interactions.
Given that  a fermion singlet cannot have direct renormalizable couplings to SM particles, one needs to add something else
and many possibilities arise.
For example one can add an inert Higgs doublet $H'$ allowing for the coupling $\psi H' L$: then the weak gauge interactions
of $H'$ would produce the `finite naturalness' bound computed in eq.\eq{MDMS}.
As we here want to explore DM systems without SM gauge interactions, we add a neutral scalar singlet $S$, such that the
Lagrangian is~\cite{fermionsinglet}
\beq \Lag = \Lag_{\rm SM} +\frac{(\partial_\mu S)^2}{2}+\bar\psi i\slashed{\partial}\psi
-\frac{m_S^2}{2}S^2-\frac{\lambda_S}{4} S^4-\lambda_{HS} S^2 |H|^2+ \frac{y }{2} S \psi\psi+\frac{M_\psi}{2}\psi\psi+{\textrm{h.c.}}
\eeq
We consider the  minimum where both the Higgs and $S$ get a vacuum expectation value:
\beq
v^2=\frac{\lambda_S m^2 + 2\lambda_{HS} m_S^2}{4(\lambda\lambda_S-\lambda_{HS}^2)},\qquad
\langle S\rangle^2=\frac{4\lambda m_S^2+2\lambda_{HS} m^2}{4(\lambda_{HS}^2-\lambda_S\lambda)}.
\eeq
The mixing among $h$ and $S$ leads to scalar mass eigenstates $S_1$ and $S_2$:
\beq
S=\cos\alpha \,S_2-\sin\alpha \,S_1\,,\qquad
h=\sin\alpha \,S_2+\cos\alpha \,S_1
\eeq
where the mixing angle $\alpha$ is
\beq
\tan 2\alpha=\frac{4\sqrt 2  \lambda_{HS}v \langle S\rangle}{\lambda v^2-2\lambda_S \langle S\rangle ^2}.
\eeq
 We identify $S_1$ with the 125 GeV Higgs field.
 The one-loop correction to the Higgs squared mass is then
\beq \delta m^2 = \frac{6 y ^2 \sin^2{\alpha}}{(4\pi)^2} M^2  (  \ln\frac{M^2}{\bar\mu^2} -\frac{1}{3})\label{eq:dmFS}
\eeq
where $M = M_\psi + y \langle S\rangle$ is the DM mass.
Its spin-independent  cross section on nucleons is given by~\cite{fermionsinglet}
\beq
\sigma_{\rm SI}=\frac{y ^2 \sin^2{2\alpha}}{8\pi}\frac{m_N^4 f^2}{v^2}\left(\frac{1}{m_1^2}-\frac{1}{m_{2}^2}\right)^2.
\eeq

%At one loop there is  the  correction to $m^2$ induced by the fermion \marco{penso sia giusto, da controllare}:
%\beq \delta m^2 =  \frac{4 y ^2}{(4\pi)^2 } M_{DM}^2 s_{\alpha}^2 (\ln\frac{M_{DM}^2}{\bar\mu^2} -1)
%\eeq
%where $M_{DM}=M_\psi+2 y  \langle S \rangle$, thus reproducing the same result as in Type-I see saw with an additional suppression factor due to mixing.

\begin{figure}[t]
\begin{center}
$$\includegraphics[width=0.45\textwidth]{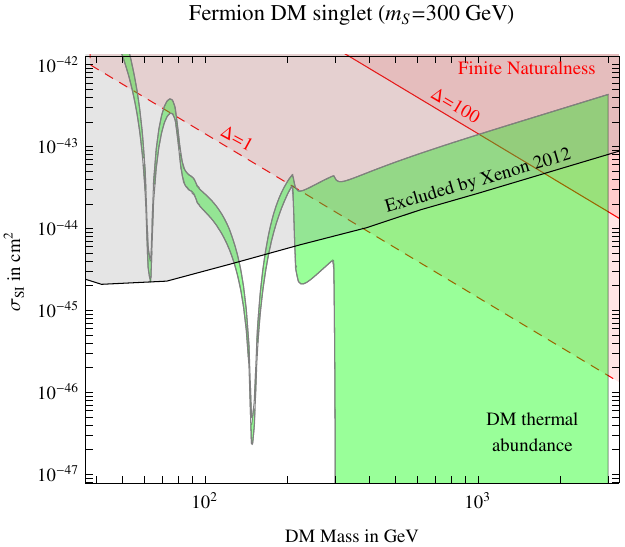}\qquad
\includegraphics[width=0.45\textwidth]{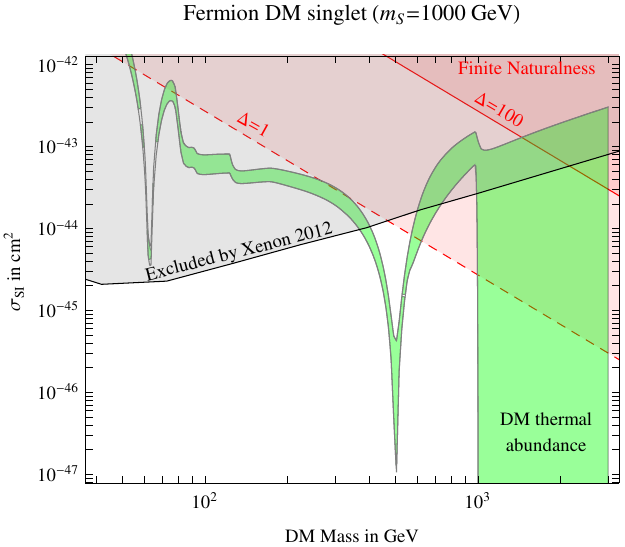}$$\vspace{-1cm}
\caption{\em  {\bf Fermion singlet DM model}.
In the ($M,\sigma_{\rm SI}$) plane we plot the region excluded by the Xenon direct search experiment
and the upper bound on the DM mass following from
finite naturalness for fine-tuning $\Delta=1$ (dashed line) and $\Delta=100$ (continuous line).
We assume $m_{S_2}=300\GeV$ (left) or $1\TeV$ (right).
This does not fix all parameters of the model
such that
the region where the DM thermal density can reproduce the cosmological DM density (green band)
becomes wide at $M>m_{S_2}$.
We demand compatibility with electroweak precision data.
\label{fig:FS}}
\end{center}
\end{figure}

The DM annihilation cross section, at tree level, is given by~\cite{fermionsinglet}
\beq\label{psipsitoX}
\sigma (\psi\psi\to X)=\sigma(\psi\psi\to {\rm SM})+\sum_{i,j=1,2}\sigma(\psi\psi\to S_iS_j).
\eeq
Both terms are $p$-wave suppressed.
The first term includes all annihilations into SM particles mediated by $s$-channel exchange of one of the two scalars $S_i$:
\beq
\sigma(\psi\psi\to {\rm SM})=v_{\rm rel} \frac{y ^2M^2 \sin^2{2\alpha} }{4}
\frac{\Gamma_{h\to {\rm SM}}(2M)}{2M}
\left|\frac{i}{4 M^2-m_1^2+im_{1}\Gamma_{S_1}}-(1\leftrightarrow 2)\right|^2
\eeq
where $v_{\rm rel}$ is the relative DM velocity,
$\Gamma_{h\to {\rm SM}}(m)$ is the width of a SM Higgs of mass $m$ and $\Gamma_{S_i}$ are the widths of the two physical scalars in the model.
This contribution vanishes in absence of mixing, namely  when $\langle S\rangle=0$.

The second term in eq.~(\ref{psipsitoX}) receives two kinds of contributions: $s$-channel exchange of the two scalars and $t$ and $u$-channel exchange of $\psi$. The former process is proportional to the quartic coupling $\lambda_{HS}$ which we assume to be small, while the latter is non vanishing also in the absence of mixing between $S$ and $h$. It is given by
\beq
\sigma(\psi\psi\to S_iS_j)=v_{\rm rel} \frac{3 y ^4 c_i^2 c_j^2  f_{ij}}{64\pi M^2(1+\delta_{ij})}
\eeq
where $c_1=\sin\alpha$, $c_2=\cos\alpha$
and
\beq
f_{ij}=\tfrac{2304M^8-1024 M^6(m_i^2+m_j^2)+32M^4(3m_i^2+m_j^2)(m_i^2+3m_j^2)+(m_i^2-m_j^2)^4}{9(4M^2-m_i^2-m_j^2)^4}\sqrt{1+\tfrac{(m_i^2-m_j^2)^2}{(2M)^4}-\tfrac{2(m_i^2+m_j^2)}{(2M)^2}}
\eeq
normalized to satisfy $f_{ij}= 1$ for $m_{i,j}\to 0$.

\bigskip

The  number of free parameters (five) of the model and the variety of possible resonant enhancements
make it difficult to analyze it.
In fig.\fig{FS} we consider the ($M,\sigma_{\rm SI}$) plane relevant for direct DM searches.
We also fix $m_{S_2}$ such that resonances arise at fixed values of the DM mass $M$, and such that
the quantum correction to the Higgs mass, eq.\eq{dmFS}, is determined in terms of
$M,\sigma_{\rm SI}$ and $m_{S_2}$.
This allows us to computed the `finite naturalness' bound, plotted in fig.\fig{FS}.
Furthermore, we plot the range (in green) where
the cosmological DM abundance can be reproduced thermally;
in view of the extra undetermined parameters such band becomes wide at $M>m_{S_2}$.
Its sharp structure corresponds to the various possible resonances.
Such band has been restricted imposing that Higgs data and electroweak precision tests are satisfied.\footnote{
The mixing between $S$ and the Higgs corrects the electroweak parameters as
\bea
\hat T&=& \frac{3g_2^2}{64\pi^2}\left[\sin^2\alpha \left(f_T\left(w_2\right)-\frac{f_T\left(z_2\right)}{c_W^2}\right)-\cos^2\alpha \left(f_T\left(w_1\right)-\frac{f_T\left(z_1\right)}{c_W^2}\right)\right],\\
\hat S&=& \frac{g_2^2}{32\pi^2}\left[\sin^2\alpha\,  f_S\left(z_2\right)-\cos^2\alpha\, f_S\left(z_1\right)\right],
\eea
where $z_i=m_i^2/m_Z^2$ and $w_i=m_i^2/m_W^2$. The functions $f_S$ and $f_T$ are given by
\bea
f_S(x)&=&\frac{1}{12}\left[-2x^2+9x+\left(x^2-6x+18\frac{x-2}{x-1}\right)x\log x+2\sqrt{x^2-4x}(x^2-4x+12)g(x)\right]
\nonumber \\
f_T(x)&=&\frac{x\log x}{x-1},\qquad
g(x)= \left\{ \begin{array}{ll}
\tan^{-1}\left(\frac{\sqrt x}{\sqrt{x-4}}\right) -\tan^{-1}\left(\frac{x-2}{\sqrt{x^2-4x}}\right) & \hbox{if $0<x<4$}\cr
\frac{1}{2}\log\frac{1}{2}\left(x-2-\sqrt{x^2-4x}\right)& \hbox{if $x>4$}
\end{array}\right. \ .
\eea}
We see that DM is constrained by `finite naturalness' to be lighter than a few TeV in the region where
direct detection experiments can test this model.

%This is shown in fig.~(\ref{fig:EWPTS}).
%For any fixed value of $\alpha\neq 0$, perturbativity of the Higgs quartic $\lambda$ give an upper bound on $m_S$. This is also shown in Fig.~(\ref{fig:EWPTS}) for two representative values of $\lambda_H$.
%For fixed value of $\lambda$ there is also a lower bound on $m_S$, Appendix~(\ref{app:singletfermion}).

%\begin{figure}[t]
%\begin{center}
%$$\includegraphics[width=0.45\textwidth]{figs/EWPTS.pdf}\qquad
%\includegraphics[width=0.45\textwidth]{figs/fermionsinglet1000.pdf}$$
%\parbox{0.45\textwidth}{
%\caption{\em 1, 2 and 3$\sigma$ contours (full, dashed and dotted respectively) from LEP-Tevatron electroweak fit. In red the $\lambda_H$ isolines.
%\label{fig:EWPTS}}}\hspace{0.06\textwidth}
%\parbox{0.45\textwidth}{
%\caption{\em  {\bf Fermion singlet DM model}.
%In the ($M,\sigma_{\rm SI}$) plane we plot the region excluded by the Xenon direct search experiment,
%the band favoured by the thermal DM abundance, the upper bound on the DM mass following from
%finite naturalness for $\Delta=1$ and $\Delta=100$. In this case $m_S=1000$ GeV while $\alpha$ is only required to be compatible with EWPT. \marco{Si capisce?}
%\label{fig:FS}}}
%\end{center}
%\end{figure}

%\subsection{The doublet singlet model}
% (see http://arxiv.org/pdf/0705.4493v3.pdf).
%It contains two left-handed fermion doublets $D$, $D^c$ of hypercharge $\pm\tfrac{1}{2}$ and a left handed singlet $S$ with a lagrangian
%
%\beq \Lag = \Lag_{\rm SM} + \lambda DHS+\lambda'D^cH^\dagger S+ M_D DD^c+\frac{M_S}{2} S^2 + {\textrm{h.c.}}\eeq

\section{Finite naturalness and axions}\label{CP}
Axions are a solution to the strong CP problem~\cite{axionmodels} that requires new physics coupled to the SM at a scale $f_a$ which is experimentally
constrained to be $f_a\circa{>}10^9\GeV$, much above the weak scale~\cite{axioncooling}.
Furthermore, axions can be DM and provide the observed DM density via the `misalignment' production mechanism provided that
$f_a \approx 10^{11}\GeV$, or higher in presence of fine-tunings in the phase of the initial axion vacuum expectation value~\cite{axionDM}.

So, the main issue is if these larger scales mean that finite naturalness is unavoidably violated.
As we will see, the answer is no.
To address this issue we separately analyze the two main classes of axion models.

\subsection{KSVZ axions}
 KSVZ axion models~\cite{KSVZ} employ new heavy quarks $\Psi$ whose mass arises only from the
vev of a complex axion scalar $A$ that spontaneously  breaks a global U(1)$_{\rm PQ}$ symmetry:
\beq \Lag = \Lag_{\rm SM} + \bar\Psi i\slashed{\partial} \Psi + |\partial_\mu A|^2+
(\lambda_{A\Psi}A \Psi_L \Psi_R +\hbox{h.c.}) +V(A) + \lambda_{AH} |AH|^2\eeq
The light axion $a$ is the phase of $A$.
Integrating out the heavy quarks $\Psi$ gives rise to the usual anomalous couplings of the light axion
suppressed by $f_a$, even if the heavy quark mass $M = \lambda_{A\Psi} \langle A \rangle = \lambda_{A\Psi} f_a e^{ia/f_a}$
is much smaller than $f_a$.

\medskip

Finite naturalness holds if the coupling $\lambda_{AH}$ is smaller than about $(4\pi v/f_a)^2$.
This coupling is unavoidably generated from RGE at two loops as $\delta \lambda_{AH} \sim g^2 \lambda_{AQ}^2/(4\pi)^4$.
This effect is equivalent to computing the two loop correction from heavy fermions $\Psi$ to the Higgs mass term,
which is precisely given by eq.\eq{dmmFermion} times the colour multiplicity of $\Psi$ (which is 3 for colour triplets).
Then, finite naturalness implies the following upper bounds on the mass of the heavy fermions
\beq M \circa{<}\sqrt{\Delta}\times \left\{ \begin{array}{ll}
0.74\TeV & \hbox{if $\Psi = Q\oplus \bar Q$}\cr
4.5\TeV & \hbox{if $\Psi = U\oplus\bar U$}\cr
9.1\TeV & \hbox{if $\Psi = D\oplus\bar D$}
\end{array}\right.
\label{eq:dmmaxion}\eeq
having assumed that they have the same quantum numbers as the SM quarks usually denoted as
$Q$, $U$ and $D$.
Furthermore, there is a 3-loop RGE effect which applies
even if  $\Psi$ has only strong interactions and no electroweak interactions:
the heavy quarks $\Psi$ interact with gluons, that interact with the top quark, that interacts with the Higgs, so that
\beq \delta m^2 \sim \frac{g_3^4 y_t^2}{(4\pi)^6} M^2 \ln\frac{M^2}{\bar\mu^2}.\eeq
This correction implies a finite naturalness bound on $M\circa{<}\sqrt{\Delta}\times10\TeV$.
With multiple heavy fermions, the bounds in\eq{dmmaxion} apply to the heaviest one.

In conclusion, the expectation from finite naturalness and KSVZ axions
is heavy coloured new fermions with a mass around the TeV scale.
In the absence of Yukawa couplings $H\psi\Psi $
(between the Higgs $H$, the SM quarks $\psi$ and the heavy quarks $\Psi$) the
latter would appear at LHC as stable heavy hadrons;
their cosmological thermal relic abundance would be suppressed by the strong annihilation cross section.
If $H\psi\Psi $ is present, $\Psi$ would behave as heavy quarks;
the interactions of such heavy quarks in the thermal plasma of the early universe would lead to a population of axions
with $T_a \approx 0.903\,{\rm K}$.
If $m_a\ll T_a$ this corresponds to extra radiation equivalent to
$\Delta N_\nu \approx 0.026$ effective neutrinos, compatibly with existing data~\cite{Planck}.

%\duccio{0.87 K for $\bar Q\oplus Q$, 0.89 K for $\bar U\oplus U$.
%%  today g_*s Å 3.91, before g_s Å 106.75, so Ta Å 2.78 K (g_*s/g_s)^(1/3)
%
%Number of relativistic degrees of freedom at recombination
%\beq
%g_*=g_\gamma+\frac{7}{8} g_\nu N^{SM}_{eff} \left(\frac{T_\nu}{T}\right)^4+\left(\frac{T_a}{T}\right)^4
%\eeq
%The shift in $N_{eff}$ due to axion is then
%\beq
%\Delta N_{eff}= \frac{8}{7 g_\nu}\left(\frac{T_a}{T_\nu}\right)^4.
%\eeq
%The ratio $T_a/T_\nu$ is obtained by
%\beq
%(g_{SM}+g_\Psi) T_a^3=(g_\gamma+g_\nu+g_{e^\pm})T_\nu^3,\quad  \Rightarrow \quad\frac{T_a}{T_\nu}=\left(\frac{g_\gamma+g_\nu+g_{e^\pm}}{g_{SM}+g_\Psi}\right)^{1/3}
%\eeq
%Since
%\beq
%g_{SM}=106.75,\, g_\gamma=2,\, g_\nu=\frac{7}{8}\times 6,\,g_{e^\pm}=\frac{7}{8}\times 4
%\eeq
%and
%\beq
%g_{\bar Q\oplus Q}=\frac{7}{8}\times 2\times 12,\, g_{\bar U\oplus U}=g_{\bar D\oplus D}=\frac{7}{8}\times 2\times 6
%\eeq
%one has
%\beq
%\Delta N_{eff}^{\bar Q\oplus Q}=8\times 10^{-3},\,\Delta N_{eff}^{\bar U\oplus U}=9\times 10^{-3}
%\eeq
%Se uso $g_\Psi=0$, $T_a=0.92 K$ e $\Delta N_{eff}=0.01$.}

\subsection{DFSZ axions}  DFSZ axion models~\cite{DFSZ} employ two Higgs doublets $H_u$ and $H_d$ and a complex  scalar $A$
without SM gauge interactions
such that the Lagrangian
\beq \Lag \supset
(  \lambda_U\,  UQH_u+ \lambda_DDQH_d + \lambda_{AH}A^2 H_u H_d  +  \hbox{h.c.}) + V(A) + V(H_u,H_d).\eeq
has a global U(1)$_{\rm PQ}$ symmetry broken by the vacuum expectation value of $A$.
Since the SM quarks $Q,U,D$ have non zero charges under the PQ symmetry, they generate the QCD-QCD-PQ anomaly.
The condition of finite naturalness arises already at tree level, $\delta m^2 \sim \lambda_{AH} f_a^2$
and demands a very small $\lambda_{AH}\circa{<} (v/f_a)^2\circa{<} 10^{-15}$.

%OR
%With $V(A) =  \frac{M_A^2}{2} |A|^2 - \frac{\lambda_A}{4} |A|^4$ one has
%$f_a = M_A/\sqrt{\lambda_A}$ and
%$m_A = M_A$: maybe a small $\lambda_A$ is also good?

%In DFSZ models $\delta m^2 \sim \lambda_{AH} M_A^2$ but the problem arises already at tree level

\section{Finite naturalness, vacuum decay and inflation}\label{vac}
Another experimental result which might suggest the presence of physics beyond the SM is the fact that
the SM potential (for the currently favored values of the Higgs and top masses) develops an
instability at field values above about $10^8\GeV$, leading to vacuum decay with a rate much longer than
the age of the universe~\cite{SMextrapolated}.

There are many ways to avoid this instability, which employ loop corrections from new
particles with sizeable  couplings to the Higgs~\cite{instabsol}.
Thereby, in the context of finite naturalness, this kind of new physics is expected to be around the weak scale.

This is however not a general conclusion.
Indeed there is one special model where the instability is avoided by a tree level effect with small couplings.
Adding to the SM a scalar singlet $S$ with interactions to the Higgs described by the potential~\cite{SS}
\beq
V=\lambda_H \left( H^\dagger H -v^2\right)^2 +\lambda_S  \left( S^\dagger S -w^2\right)^2
+2\lambda_{HS}  \left( H^\dagger H -v^2\right)  \left( S^\dagger S -w^2\right)
\label{pot}
\eeq
the low-energy Higgs quartic coupling is given by
$\lambda =\lambda_H-{\lambda_{HS}^2}/{\lambda_S}$ at tree level.
This model allows to stabilize the SM vacuum compatibly with `finite naturalness' even if the singlet is much above the weak scale, provided that
the couplings $\lambda_{HS}$ and $\lambda_S$ are small.
A singlet with this kind of couplings is present within an attempt of deriving the SM from the framework of non commutative geometry~\cite{Connes}.

\bigskip

Finally, observations of cosmological inhomogeneities suggest that the full theory incorporates some mechanism for inflation.
At the moment the connection with the SM is unknown, even at a speculative level.
A successful inflaton must have a flat potential, which is difficult to achieve in models;
at quantum level flatness usually demands small couplings of the inflaton to SM particles.
An inflaton decoupled from the SM would satisfy `finite naturalness'.
A free scalar $S$ with mass $M\approx 10^{13}\GeV$ is the simplest inflaton candidate;
it satisfies finite naturalness provided that its couplings to the Higgs
$\lambda_{HS}$ is smaller than about $10^{-10}$.
It is interesting to notice that this roughly is the maximal mass compatible with `finite naturalness':
%gravity generates a minimal $\lambda_{HS}\sim M^2/M_{\rm Pl}^2\sim 10^{-12}$
at three loops gravity\footnote{We thank A. Arvinataki, S. Dimopoulos and S. Dubovsky
for having noticed and pointed to us such effect.}
gives a finie correction to the Higgs mass, $\delta m^2 \sim y_t^2 M^6/M_{\rm Pl}^4(4\pi)^6$.
%\xxx{Maybe this applies?
%N.~Arkani-Hamed, L.~Motl, A.~Nicolis and C.~Vafa,
%  %``The String landscape, black holes and gravity as the weakest force,''
%  JHEP {\bf 0706} (2007) 060
%  [hep-th/0601001].}

In general, inflation can occur at lower scales.
A model of inflation at the weak scale based on a singlet $S$
was proposed in~\cite{EWinflation} and anyhow needs a small coupling to the Higgs to preserve flatness.

\section{Conclusions}\label{concl}
Experiments  are clarifying whether
the smallness of the Higgs mass respects naturalness or not.
Understanding the origin of the weak scale is particularly
important from a meta-physical point of view,
because here is where two different  views of the universe clash:
do we live in a `good' supersymmetric universe with no tunings,
or in a `bad' multiverse
where huge tunings make some of its regions anthropically observable?

We here explored the `ugly' third possibility: that the usual formulation of naturalness~\cite{thooft}  must be modified by ignoring
the uncomputable quadratically divergent corrections to the squared Higgs mass, and keeping only the finite computable corrections.

\smallskip

%Let us compare the usual formulation of naturalness to the weaker condition of `finite naturalness' from a practical point of view.
The usual naturalness is not satisfied by the SM, and
demands supersymmetry, or some other big modification of particle physics.
Such new physics must be around the weak scale
or at most a one-loop factor above it, as computed in eq.\eq{stop} for the coloured stops.

\smallskip

The modified `finite naturalness' is satisfied by the SM
for the measured values of its parameters, as we have shown by computing
the Higgs mass parameter  with one-loop accuracy.

However, experimental data demand new physics.
The consequent extension of the Standard Model can satisfy `finite naturalness'
provided that such new physics is not much above the weak scale.
The connection with the weak scale often arises at two-loops, such that
`finite naturalness' is not yet challenged by negative results of experimental searches.
We find that:
%Next, we classified models of new physics suggested or demanded by experimental data from the point of view of `finite naturalness':
\begin{itemize}
\item {\bf Neutrino masses} can be mediated at tree level by three kinds of see-saw models, that employ new particles
with mass $M$.  We find the following finite naturalness bounds:
\beq M \circa{<}\left\{\begin{array}{ll}
0.7~10^7\GeV\times\sqrt[3]{\Delta}  & \hbox{type I see-saw model},\\
200\GeV \times\sqrt{\Delta} & \hbox{type II see-saw model},\\
940\GeV \times\sqrt{\Delta} & \hbox{type III see-saw model},
\end{array}\right.\eeq
where $\Delta\sim 1$ is a fine-tuning factor.
Thermal leptogenesis seems possible only in the first case.

\item {\bf Dark Matter} might be a new particle.
Assuming that DM has weak interactions, we computed the `finite naturalness' bound on its mass $M$ summarizing
the results in table~\ref{tab:MDM}.
The weakest bound, $M\circa{<}1.9\TeV\times\sqrt{\Delta}$ is obtained for a fermion doublet.
Stronger bounds are obtained for scalar DM
(e.g.\ $M\circa{<}0.4\TeV\times\sqrt{\Delta}$ for a scalar doublet)
and for bigger $\SU(2)_L$ multiplets (e.g.\ $M\circa{<}1.0\TeV\times\sqrt{\Delta}$ for a fermion triplet
and $M\circa{<}0.4\TeV\times\sqrt{\Delta}$ for a fermion quintuplet).
We also considered models where DM has no SM gauge interactions and couples to the Higgs doublet,
finding bounds at the TeV level under the assumption that DM is a thermal relic and/or that
direct detection is detectable.

\item {\bf Axions} can solve the strong CP problem and can be DM provided that they are associated
with a new high scale, $f_a \sim 10^{11}\GeV$.  We have shown that axion models can be compatible with `finite naturalness';
within KSVZ axion moles compatibility implies new quarks at the TeV scale.

\item {\bf Vacuum decay and inflation} do not lead to concrete restrictions.
\end{itemize}
In conclusion, extensions of the SM motivated by observed phenomena can be compatible with `finite naturalness',
and this often demands new particles not much above the weak scale and potentially accessible to colliders.

The smallness of the cosmological constant seems to violate both versions of naturalness.

\small

\paragraph{Acknowledgements}
We thank Savas Dimopoulos, Gian Giudice and Mikhail Shaposhnikov for discussions and N. Nagata 
for a comparison about the last column of table~\ref{tab:MDM}.
This work was supported by the ESF grants  8090, 8499, MTT8 and by SF0690030s09 project. The work of MF was supported in part by the NSF grant PHY-0757868.

\footnotesize
\begin{multicols}{2}

\end{multicols}

\end{document}